\documentclass[aps,prb,twocolumn,superscriptaddress,longbibliography]{revtex4-2}
\usepackage{amsmath,amssymb}
\usepackage[pdftex]{hyperref,graphicx}
\hypersetup{colorlinks = true, urlcolor = blue, linkcolor = blue, citecolor = blue}
\usepackage{physics}
\usepackage{xcolor}
\usepackage{bm}
\usepackage[justification=raggedright,singlelinecheck=false]{caption}
\usepackage{subcaption}
\newcommand{\comment}[1]{}

\newcommand{\hl}[1]{{\leavevmode{#1}}}

\newcommand{\HPcom}[1]{{\leavevmode\color{cyan}{[HP: #1]}}}
\newcommand{\JTadd}[1]{{\leavevmode{#1}}}

\newcommand{\cmVs}{\text{cm}^2/\text{Vs}}

\begin{document}
\title{Ballistic transport in 1D Rashba systems in the context of Majorana nanowires}
\author{Haining Pan}
\affiliation{Department of Physics and Astronomy, Center for Materials Theory, Rutgers University, Piscataway, NJ 08854 USA}
\author{Jacob R. Taylor}

\author{Jay D. Sau}
\author{Sankar Das Sarma}
\affiliation{Condensed Matter Theory Center and Joint Quantum Institute, Department of Physics, University of Maryland, College Park, MD 20742, USA}

\begin{abstract}
Recent work on Majorana-bound states in semiconductor-superconductor hybrid structures has elucidated the key role of unintentional (and unknown) disorder (producing low-energy Andreev-bound states) in the system, which is detrimental to the emergence of Majorana-carrying topological superconductivity artificially engineered through the combination of superconductivity, Zeeman spin splitting, and Rashba spin-orbit coupling. In particular, the disorder must be smaller than the superconducting gap for the appearance of Majorana modes, but the disorder-induced appearance of subgap Andreev-bound states suppresses the Majorana modes. We theoretically investigate, as a function of disorder, the normal state ballistic transport properties of nanowires with and without superconductors in order to provide guidance on how to experimentally estimate the level of disorder. Experimentally, the superconductivity is suppressed simply by rotating the magnetic field appropriately, so both physics can be studied in the same set-up. In particular, the presence of spin-orbit coupling and Zeeman splitting produces a helical gap in the 1D electronic band structure, which should have clear signatures in ballistic transport unless these signatures are suppressed by disorder and/or Fabry-P\'erot resonances associated with the finite wire sizes. Our work provides a benchmarking of when and what signatures of the putative helical gap (which is essential for the emergence of Majorana modes by leading to a single Fermi surface) could manifest in realistic nanowires. We also provide results in the presence of a superconductor (with the Zeeman field oriented such that the nanowire remains normal) to make contact with a recent experiment. A main conclusion is that the system could actually possess a Majorana-carrying helical gap even if the normal state ballistic transport results are ambiguous or agnostic about such a gap.

\end{abstract}

\maketitle

\section{Introduction}

\begin{figure*}[ht]
    \centering
    \includegraphics[width=6.8in]{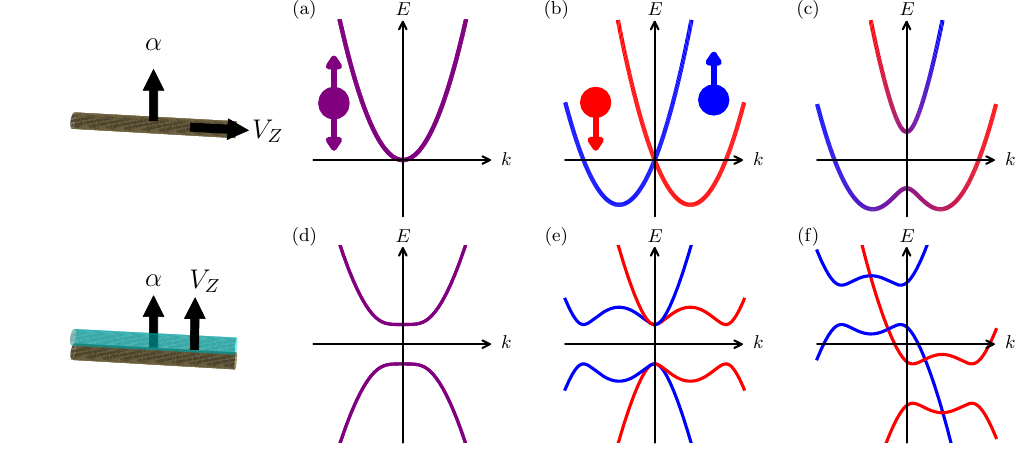}
    \caption{
        Top row: {Energy spectrum for} semiconductor nanowire (green) with perpendicular Rashba-type spin-orbit coupling $\alpha$ and Zeeman field $V_Z$ for (a) zero $\alpha$ and $V_Z$; (b) finite $\alpha$ and zero $V_Z$; (c) finite (and sufficiently large) $\alpha$ and $V_Z$. The purple implies the spin degenerate states, while the blue and red correspond to the $s_x$ spin up and down states, while the gradient color indicates the spin hybridization state with an approximate spin direction.
        Bottom row: {Energy spectrum for} semiconductor (green)--superconductor (cyan) hybrid nanowire with parallel Rashba type spin-orbit coupling $\alpha$ and Zeeman field $V_Z$ for (d) zero $\alpha$ and $V_Z$; (e) finite $\alpha$ and zero $V_Z$; (f) finite $\alpha$ and $V_Z$.
        Both (c) and (f) show helical gaps. 
        }
    \label{fig:1}
\end{figure*}

In 2009--10, a specific theoretical proposal was made to artificially create a 2D topological superconductor, which could host non-Abelian Majorana bound states, using a hybrid semiconductor (SM)-superconductor (SC) structure where the SC induces proximity-induced superconducting gap in the SM, which, in the presence of spin-orbit coupling (SOC) and spin splitting in the SM, could lead to effectively spinless $p$-wave topological superconductivity.~\cite{sau2010generic,sau2010nonabelian} \comment{[cite Sau PRL, and Sau PRB 2010]}  The experimental focus quickly shifted to a 1D size-quantized version of this SM-SC proposal (with identical physics to the 2D case), where a SM nanowire with strong SOC (e.g., InAs, InSb) is proximitized by a SC film (e.g., Al, Nb), and the spin splitting is created by applying a Zeeman field along the wire.~\cite{lutchyn2010majorana,oreg2010helical}\comment{  [cite Lutchyn and Oreg PRLs 2010]}  The chiral Majorana modes of the 2D version become end-quantized in the `helical' nanowire, which is reminiscent of a toy model introduced earlier by Kitaev assuming spinless $p$-wave SC in a 1D lattice chain.~\cite{kitaev2001unpaired}\comment{ [cite Kitaev]}  
This SM-SC nanowire system has been extensively studied both experimentally and theoretically~\cite{nayak2008nonabelian,wilczek2012quantum,leijnse2012introduction,alicea2012new,stanescu2013majorana,beenakker2013search,sarma2015majorana,elliott2015colloquium,sato2016majorana,aguado2017majorana,sato2017topological,lutchyn2018majorana, flensberg2021engineered,sau2021chapter,laubscher2021majorana,marra2022majorana,dassarma2023search, amundsen2024colloquium,yazdani2023hunting,kouwenhoven2025perspective,sarma2025rashba}\comment{[cite all the reviews we cite in our recent Rashba review as well as the Rashba review itself]}, and forms the qubit platform for Microsoft Corporation in their efforts to build a topological quantum computer by fusing/braiding the end-localized Majorana modes.~\cite{microsoftquantum2023inasal,aghaee2025interferometric, aasen2025roadmap}\comment{[cite all the recent MSFT paper as cited in our Rashba review ]} It is well-established that the main impediment in the nanowire Majorana development is the disorder arising from unintentional random impurities which are invariably present in the SM-SC hybrid structures, both inside the SM material, and perhaps more importantly, at the SM-SC interface and other interfaces present in the rather complicated hybrid structure.~\cite{sau2012experimental,dassarma2023search,dassarma2023search, brouwer2011probability, liu2012zerobias, liu2012zerobias, akhmerov2011quantized, sau2013density, takei2013soft, adagideli2014effects, pekerten2017disorderinduced,brzezicki2017driving, barmanray2021symmetrybreaking, pan2020physical, ahn2021estimating, woods2021chargeimpurity, dassarma2021disorderinduced, dassarma2023density, dassarma2023spectral, pan2024disordered, taylor2024machine, fulga2011scattering, hegde2016majorana, penaranda2018quantifying, yu2021nonmajorana}\comment{[cite all the disorder papers from our Rashba review]}  Disorder strength must be smaller than the SC gap for topological superconductivity and Majorana states to emerge, and the topological gap depends on the strength of SOC, which is also not known in the hybrid structures. This has remained the most serious problem, seriously hindering progress   \cite{dassarma2023search,kouwenhoven2025perspective}. In particular, all early experiments (up to 2020) only observed disorder effects, which, with fine-tuning, could occasionally mimic some of the Majorana signatures, and hence there were (unintentional) incorrect claims of observing `Majorana signatures' in the SM-SC structures.~\cite{mourik2012signatures,das2012zerobias,deng2012anomalous,churchill2013superconductornanowire,nichele2017scaling,pan2020physical,dassarma2021disorderinduced,zhang2021large,dassarma2023search} \comment{[cite Pan-Das Sarma and Das Sarma-Pan, and Das Sarma Nature Physics as well as Zhang et al arXiv which replaces the 2018 Zhang Nature]}  In the context of developing some direct quantitative understanding of the disorder content in the experimental Majorana nanowires, it is highly desirable to estimate the amount of disorder in the nanowire itself without the complications of the superconductor and topology. This can, in principle, be done in two alternative (but complementary) ways. The ballistic conductance in the nanowire itself could be studied in the presence of Zeeman splitting and SOC, but without any SC, as a function of disorder. Also, the ballistic conductance can be studied by adding the SC, but with the magnetic field oriented transverse to the nanowire along the direction of spin-orbit coupling,
so that the system does not enter the topological phase. Both situations would provide direct quantitative information about the disorder content in the SC-SM nanowire structure as well as disorder impact on the helical gap which should exist in a 1D Rashba system in the presence of spin-orbit coupling and Zeeman splitting (see Fig. \ref{fig:1}). This is what we investigate theoretically in this work for realistic SM-SC systems, using the parameters for the recent experimental nanowires in the InAs/Al hybrid structures.~\cite{microsoftquantum2023inasal}

We calculate the electronic ballistic conductance in nanowires (nominally for the InAs/Al hybrid SM-SC structures) as used in \cite{microsoftquantum2023inasal} both with and without the SC, but in the presence of Zeeman and spin-orbit effects. The system without the SC is simply a 1D wire in the strict one-subband limit, where the physics is well-understood in the absence of SOC and Zeeman splitting in the pristine 1D ballistic limit, manifesting $2e^2/h$ conductance quantization, which is generic in the 1D limit.~\cite{he1989quantum,he1993quantum,vanwees1988quantized,wharam1988onedimensional} \comment{[cite ~https://journals.aps.org/prb/abstract/10.1103/PhysRevB.40.3379;~https://journals.aps.org/prb/abstract/10.1103/PhysRevB.48.4629; ~https://journals.aps.org/prl/abstract/10.1103/PhysRevLett.60.848; ~https://iopscience.iop.org/article/10.1088/0022-3719/21/8/002]}  In the presence of Zeeman splitting, the spin split subbands produce $e^2/h$ conductance quantization because of the lifting of the spin degeneracy.~\cite{vanwees1988quantized}\comment{ [https://journals.aps.org/prb/abstract/10.1103/PhysRevB.38.3625]} But the presence of SOC should, in principle, drastically change the situation, if the magnetic field is oriented along the wire, since the 1D band structure develops a helical gap under these conditions because of the presence of both spin splitting and SOC (Fig.~\ref{fig:1}) {when SOC is large enough}. The key aspect of the helical gap is that it separates two spin-degenerate regimes with an intermediate (effectively) single-spin non-degenerate regime, thus implying that the quantized 1D conductance should show a re-entrant behavior with increasing chemical potential, going from $2e^2/h$ back again to $2e^2/h$ with a region of $e^2/h$ quantization in between. We emphasize that this scenario of a helical gap induced re-entrance of conductance quantization $2e^2/h-e^2/h-2e^2/h$\comment{Jacob: The comment said this, please check this equation.} with increasing chemical potential happens only in the ideal disorder-free ballistic limit, with the size of  re-entrant regime in chemical potential  being proportional to the SOC strength (see Fig. \ref{fig:1}). This re-entrant ballistic conductance behavior depends quantitatively on the SOC strength and the Zeeman splitting energy, but, in principle, it should always be present because of the existence of the helical gap.~\cite{pershin2004effect} \comment{[cite: https://journals.aps.org/prb/abstract/10.1103/PhysRevB.69.121306]}  By contrast, if the magnetic field is oriented perpendicular to the wire, then there is no helical gap, and the system is just spin-split because of Zeeman effect, and the conductance is quantized in units of $e^2/h$ with no reentrance. We consider both of these situations separately, with magnetic field oriented along and normal to the wire, where the first case explores the helical gap effects in conductance without any superconductor and the second case corresponding to a recent experiment~\cite{microsoftquantum2023inasal} \comment{([cite MSFT PRB here]} where the magnetic field is applied in the SM-SC hybrid system to suppress the superconductivity through Pauli blockade. We refer to the first situation (magnetic field along the wire) as `without superconductors' and the second situation (where the magnetic field is oriented normal to the wire) as `with superconductors' (because the recent experiment is done in the presence of the SC in the SM-SC nanowire with the magnetic field oriented along the SOC direction). Figure~\ref{fig:1} provides detailed schematics for the nanowire band structures in both situations.

Since the existence of a helical pair of Fermi points signified by an $e^2/h$ conductance is essential to the emergence of topological superconductivity and the associated localized Majorana zero modes in nanowire SM-SC hybrid platforms ~\cite{sau2010generic,sau2010nonabelian,lutchyn2010majorana,oreg2010helical}\comment{[cite Sau 2010 PRL and PRB, Lutchyn and Oreg 2010 PRL]}, our goal in the current work is to look for signatures of 
an $e^2/h$ plateau through measurements of ballistic conductance under realistic conditions of finite disorder. We mention that conductance measurements in InSb ~\cite{kammhuber2016conductance,fadaly2017observation}
\comment{[ ~https://doi.org/10.1021/acs.nanolett.6b00051; ~https://doi.org/10.1021/acs.nanolett.7b00797]} and PbTe~\cite{wang2023ballistic}\comment{ [ https://doi.org/10.1021/acs.nanolett.3c03604]} nanowires in the context of SM-SC Majorana platforms have reported conductance quantization in units of both $e^2/h$ and $2e^2/h$ in the presence and absence of an applied magnetic field, and signatures consistent with helical states/gaps have been reported in nanowire quantum point contacts~\cite{kammhuber2017conductance,heedt2017signatures}, but the putative re-entrant helical behavior has not been reported. This is problematic because the helical gap, and not the spin splitting by the Zeeman field by itself,  is the key to providing an estimate of the spin-orbit coupling that determines the topological superconducting gap. However, the manifestation of a re-entrant helical gap in the quantization of the ballistic conductance may not be necessary for a topological superconducting phase. Topological superconductivity in systems with weak spin-orbit coupling appears without the appearance of the $2e^2/h$ conductance at low densities even though the resulting topological gap is generally quite small in such a situation. 
Our current work, presented as a function of disorder strength, provides guidelines for future experiments looking for the manifestation of the helical gap in ballistic 1D conductance. Our results, where the magnetic field is oriented normal to the wire (i.e., `with superconductors' results), so that the Zeeman field and the effective SO field are oriented parallel to each other, are to be compared with the recent measurements from Microsoft where such data are presented for the InAs/Al electron hybrid system. So, the current work has dual complementary goals: (1) provide signatures of the helical gap in the ballistic transport of realistic SM nanowires used in the Majorana experiments (Sec.~\ref{sec:SM}); (2) provide theoretical results for the transport results presented in the experiment for estimating the electron mean free path (Sec.~\ref{sec:SC}). \hl{Specifically, Sec.~\ref{sec:SM} focuses on how Fabry-P\'erot resonances and disorder obscure helical-gap conductance signatures (Figs.~\ref{fig:alpha0.5Vz0.2} and~\ref{fig:alpha0.1Vz0.2}), while Sec.~\ref{sec:SC} presents the quantitative comparison to the recent multi-terminal experiment and the resulting disorder estimate (Figs.~\ref{fig:cond}--\ref{fig:errors}).} We mention that the two sections are independent of each other and can be followed without consulting each other since the physics and the system itself are very different (except that both discuss ballistic conductance in normal 1D wires in the presence of SO coupling and Zeeman splitting, but using distinct configurations with different physics). We use parameters corresponding to the InAs/Al Majorana systems, and the main goal is to elucidate the effect of disorder on the ballistic conductance. We assume noninteracting electrons with parabolic single bands (which are degenerate in the absence of Zeeman splitting) using InAs conduction band parameters. The noninteracting approximation is excellent for InAs since the relevant $r_s$ parameter for the Coulomb interaction is very small because of the small effective mass of InAs conduction band.

The rest of this article is organized as follows.  
In Sec.~\ref{sec:SM}, we present and discuss results for a single nanowire without any superconductor for the situation where the magnetic field is oriented along the wire so that a helical gap exists.  
In Sec.~\ref{sec:SC}, following the transport measurement configuration in the experiment~\cite{microsoftquantum2023inasal}\comment{ [cite MFT PRB]}, we use the composite SM-SC system assuming the magnetic field to be normal to the wire (but along the SC plane) so that there is spin splitting, but no helical gap.  
We conclude in Sec.~\ref{sec:conclusion}  with a summary and a discussion of open questions.

\section{Normal nanowire with magnetic field and spin-orbit coupling}\label{sec:SM}
\subsection{Theory}
In this section, we first discuss the minimal example (Fig.~\ref{fig:1}(a), (b), (c))--- the 1D single-band normal nanowire in the presence of finite magnetic field and a Rashba-type spin-orbit coupling with no superconductor, where the Hamiltonian is given by~\cite{he1989quantum,he1993quantum,pershin2004effect,rainis2014conductance} $\hat{H}_{\text{NW}} = \sum_{s,s'}\int_{-\pi}^{\pi} \frac{d k}{2\pi}  \left[ H_{\text{NW}}(k) \right]_{s,s'} \psi_{s}^\dag(k)\psi_{s'}(k)$ and
\begin{equation}\label{eq:HNW}
    H_{\text{NW}}(k) = \frac{\hbar^2k^2}{2m^*} + \alpha \sigma_y k -\mu + V_Z \sigma_x.
\end{equation}
Here, $\psi_{s}^\dag(k)$ ($\psi_{s}(k)$) is the creation (annihilation) operator of an electron with spin $s=\uparrow,\downarrow$ and momentum $k$, $\sigma_{x,y,z}$ are the Pauli matrices, $m^*$ is the effective mass, and $\alpha$ is the Rashba spin-orbit coupling along the $y$ direction, $\mu$ is the chemical potential of the single band, $V_Z$ is the Zeeman energy along the $x$ direction.
In the following results, we consider the parameters in InAs nanowires $m^*=0.023m_e$.
We set $\mu=0$ and explore a small $\alpha=0.1$ eV\AA{}, and a large $\alpha=0.5$ eV\AA{} with a fixed $V_Z=0.2$ meV. We note that Eq. (1) defines the pristine non-SC part of the minimal Hamiltonian in the SM-SC Majorana nanowire platform \cite{microsoftquantum2023inasal}.

The exact diagonalization of Eq.~\eqref{eq:HNW} only provides the band structure showing the helical gap. To further demonstrate the quantization of the conductance inside the helical gap, we truncate the wire to a finite length and recast Eq.~\eqref{eq:HNW} from momentum space into real space by replacing $k\rightarrow-i\partial_x$.
To measure the conductance, we attach two normal leads to both ends of the wire, whose Hamiltonians are identical to Eq.~\eqref{eq:HNW}, but with a separate chemical potential $\mu_{L}$ ($\mu_{R}$) for the left (right) lead. 
Experimentally, the lead chemical potential should be larger than the chemical potential in the wire to ensure the propagating modes.
Here, we will study the effect of the chemical potential in the leads and therefore vary it from 0 to 10 meV. As we will show in the next section, the chemical potential of the lead strongly affects the transport signatures within the helical gap in terms of the oscillation. (One potential problem in comparing theory and experiment quantitatively is that the relative chemical potentials of the leads with the nanowire may be unknown because of the contact complications at the lead-nanowire junctions.)
  
We further discretize the continuum Hamiltonian into a tight-binding model by treating the $\partial_x $ as finite difference with a fictitious lattice constant $a=10$ nm under the total wire length of $L=1$ micron. (The results are independent of the discretization as long as $a$ is small enough, which we ensure numerically.)
The conductance is obtained by the Landauer-B\"uttiker formula from the s-matrix, where the actual calculation is performed using the efficient Python package $\texttt{KWANT}$~\cite{groth2014kwant}.

In reality, the nanowires are not pristine, and the disorder coming from the inhomogeneous electrostatic potential~\cite{woods2021chargeimpurity,ahn2021estimating}
leads to a typical mobility of 
$\sim1-10\times 10^4 \cmVs$ in the 2D InAs nanostructures~\cite{ pauka2020repairing,beznasyuk2022doubling}.
Therefore, we model the disorder in the nanowire as a random potential $V(x)$ added to the Hamiltonian Eq.~\eqref{eq:HNW}, following the uncorrelated Gaussian distribution, i.e., $\expval{V(x)V(x')}=\sigma_\mu^2\delta(x-x')a$.
In the following results, we vary the disorder strength from $\sigma_\mu=0$ to $0.3$ meV. Note that our discretization ensures a disorder correlation length of $10$ nm, which is reasonably realistic for the experimental system although the precise disorder correlation length is unknown quantitatively.

\subsection{Results}
In this section, we present the numerical results showing the band structure as a function of $k$ (black curves), overlaid with the differential tunneling conductance (red curves) from one lead to the other (We ensure that the tunneling conductance is symmetric bidirectionally) in Fig.~\ref{fig:alpha0.5Vz0.2} and Fig.~\ref{fig:alpha0.1Vz0.2} for the large $\alpha=0.5$ eV\AA{} and small $\alpha=0.1$ eV\AA{}, respectively.
In each panel, the tunneling conductance $G$ shown on the top axis is computed by setting the bias voltage to be the energy $E$ in the left axis.

\subsubsection{Large $\alpha=0.5$ eV\AA{} }
\begin{figure*}[ht]
    \centering
    \includegraphics[width=6.8in]{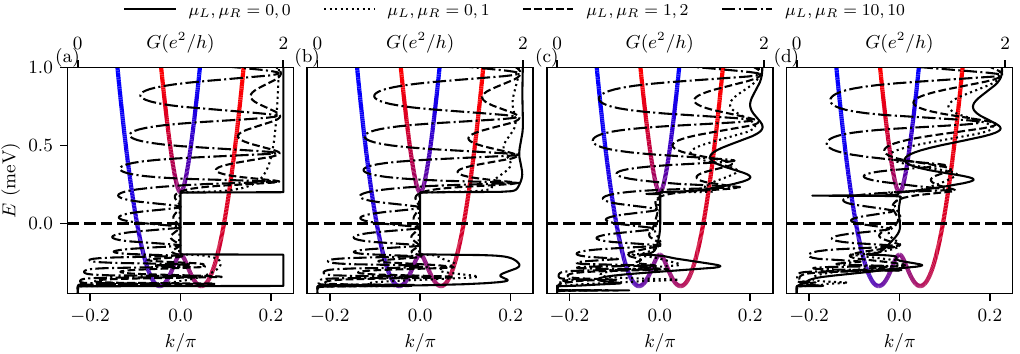}
    \caption{Tunneling conductance with various lead chemical potentials (black curves) and the band structure for approximate $s_x$ up and down (blue and red curves) in a InAs nanowire with a large spin-orbit coupling $\alpha=0.5$ eV\AA{}, wire chemical potential $\mu=0$ (horizontal dashed line), and a Zeeman field $V_Z=0.2$ meV corresponding to Fig.~\ref{fig:1}(c), showing a helical gap and the re-entrant of the $2e^2/h$ conductance from $e^2/h$.
    The disorder strength increases from left to right: pristine wire in (a); $\sigma_\mu=0.05$ meV in (b); $\sigma_\mu=0.2$ meV in (c); $\sigma_\mu=0.3$ meV in (d). The conductance is for a single disorder realization without any ensemble averaging.
    }
    \label{fig:alpha0.5Vz0.2}
\end{figure*}

To set up a baseline, we show the helical gap in the first column of Fig.~\ref{fig:alpha0.5Vz0.2}(a) with a large $\alpha$ in the absence of disorder.
We start with the lead chemical potential $\mu_{L}=\mu_{R}=0$ (solid curves), which demonstrates a helical gap between $-0.25$ and $0.25$ meV with the conductance being $e^2/h$, and a re-entrant of the conductance back to $2e^2/h$ from $e^2/h$. 
In Fig.~\ref{fig:alpha0.5Vz0.2}(a) dotted line, we increase the chemical potential on one end from 0 to 1 meV, where we find the quantization becomes less ballistic. This is analogous to Fabry-P\'erot (FP) cavity, where one side is fully transparent and the other side is reflecting.
As we keep increasing the chemical potential on both leads, as shown in the dashed and dotted dashed curve in Fig.~\ref{fig:alpha0.5Vz0.2}, we start to see the conductance oscillations, which are similar to the oscillations observed in a FP cavity, and the oscillation amplitude depends on the ``transparency'' of both leads.
The larger the difference in chemical potential between the lead and the wire leads to the higher reflection at the interface, which further enhances finesse in the conductance oscillations, in analogy to the finesse defined in the FP cavity, $\mathcal{F}=\frac{\pi(r_L r_R)^{1/4}}{1-\sqrt{r_Lr_R}}$, where $r_L$ and $r_R$ are the reflection coefficients on both sides of the cavity.  These FP resonances are intrinsic to 1D ballistic conductance, and should manifest in pristine systems generically except for some fine-tuned nongeneric values of chemical potentials.  In fact, the FP resonances should typically mask the 1D conductance quantization manifestation.

From Fig.~\ref{fig:alpha0.5Vz0.2}(a) to (d), we gradually increase the disorder strength from a pristine wire to 0.05 meV, 0.2 meV, and finally to $0.3$ meV.
As we increase the disorder strength, we find that the conductance oscillations become more pronounced and develop a richer structure.
The disorder effectively introduces additional scattering centers, which can enhance the interference effects responsible for the oscillations with shorter ``cavity'' length scales. This behavior is expected as the individual impurities introduce their own interference structures in the electron waves, producing random resonances.

\subsubsection{Small $\alpha=0.1$ eV\AA{} }
\begin{figure*}[ht]
    \centering
    \includegraphics[width=6.8in]{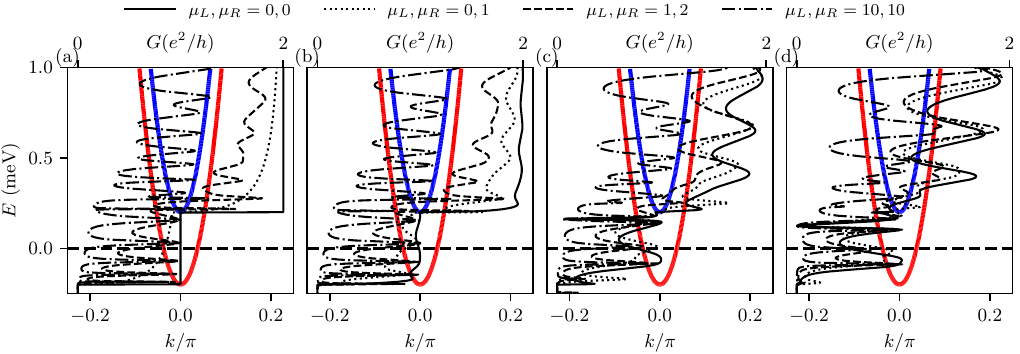}
    \caption{Tunneling conductance with various lead chemical potentials (black curves) and the band structure for approximate $s_y$ up and down (blue and red curves) in a InAs nanowire with a small spin-orbit coupling $\alpha=0.1$ eV\AA{}, wire chemical potential $\mu=0$ (horizontal dashed line), and a Zeeman field $V_Z=0.2$ meV, showing no helical gap nor the re-entrant of the quantized conductance.
    The disorder strength increases from left to right: pristine wire in (a); $\sigma_\mu=0.05$ meV in (b); $\sigma_\mu=0.2$ meV in (c); $\sigma_\mu=0.3$ meV in (d). The conductance is for a single disorder realization without any ensemble averaging.
        }
    \label{fig:alpha0.1Vz0.2}
\end{figure*}

In contrast to the large $\alpha=0.5$ eV\AA{} case, we show the small $\alpha=0.1$ eV\AA{} in Fig.~\ref{fig:alpha0.1Vz0.2}. 
Unlike the large $\alpha$ case, the small $\alpha$ case does not manifest the helical gap nor the re-entrant of the $2e^2/h$ from $e^2/h$. This is because the small $\alpha=0.1$ eV\AA{} is smaller than the critical SOC strength $\alpha_c = \sqrt{\frac{V_Z}{m^*}}\approx 0.31$ eV\AA{} for InAs necessary for re-entrance. Nevertheless, we still see the conductance oscillations as we increase the lead chemical potential.
The conductance oscillation could further be enhanced or destroyed by the disorder, as the disorder effect is complicated and depends on the specific disorder realization.



In Appendix~\ref{app:InAs_no_field_no_SOC}, we also show the band structure and conductance without a magnetic field and spin-orbit coupling. We find that the helical gap is absent, similar to the small $\alpha$ case. 
\section{SUPERCONDUCTING NANOWIRE with magnetic field and spin-orbit coupling}\label{sec:SC}

In this section we discuss transport measurements through non-local quasiparticle transport~\cite{rosdahl2018andreev} between 
the ends of the Majorana nanowire in the SM-SC Majorana nanowire structure. The advantage of this approach is that these results account for possible screening 
of impurities by the superconductor. However, this advantage comes at the cost of the superconductor screening making the transport results complicated to interpret. In fact, the transport in the conventional topological regime of the Majorana nanowire~\cite{lutchyn2010majorana,oreg2010helical}
vanishes at zero energy because of the gap of \hl{the} superconductor. This issue can be alleviated by aligning the magnetic field to be parallel 
to the spin-orbit field~\cite{microsoftquantum2023inasal} which would suppress the SC gap, but keep the parent superconductor physically present in the structure. For the case of Rashba spin-orbit coupling it means the magnetic field is oriented in the 
plane of the superconductor while being perpendicular to the nanowire. 
In this case, the Zeeman splitting does not open a topological gap, and the semiconductor nanowire remains gapless~\cite{sau2010nonabelian}. 
We model the 1D semiconductor Majorana nanowire with a superconductor and a parallel Zeeman splitting using a Bogoliubov-de–de Gennes Hamiltonian~\cite{lutchyn2010majorana}:
\begin{multline}
H = \left(-\frac{\hbar^2}{2m^*}\partial_x^2 - i\alpha\partial_x\sigma_y - \mu + V_{\text{dis}}(x)\right)\tau_z \\
+ \frac{1}{2}g\mu_B B \sigma_y + \Sigma(\omega)
\end{multline}
where the self-energy from the proximitized superconductor is
\[
\Sigma(\omega) = -\gamma \frac{\omega + \Delta_0 \tau_x}{\sqrt{\Delta_0^2 - \omega^2}}
\] as in Ref.~\cite{sau2010robustness}. Here, $\sigma_{i}$ and $\tau_{i}$ are Pauli matrices for spin and particle-hole degrees of freedom, respectively, and the Hamiltonian is written in the Nambu basis $\psi(x) = (u_\uparrow(x), u_\downarrow(x), v_\downarrow(x), {-}v_\uparrow(x))^T$. The frequency $\omega$ corresponds to the energy of the Bogoliubov quasiparticle. 

We compute the transport properties of the Majorana nanowire from the scattering matrix using the  Blonder-Tinkham-Klapwijk formalism~\cite{blonder1982transition}. The scattering matrix, in turn, is obtained from a discretized version of the Hamiltonian $H$ using  KWANT~\cite{groth2014kwant}. This model Hamiltonian quantitatively reproduces experimental transport features when parameters are appropriately fitted~\cite{dassarma2023spectral,pan2024disordered}. In particular, we use the following values: effective mass $m^* = 0.03\,m_e$, superconducting pairing potential $\gamma = 0.15\,\mathrm{meV}$, g-factor $g = 25$, superconducting gap $\Delta_0 = 0.12\,\mathrm{meV}$, spin-orbit coupling $\alpha = 8\,\mathrm{meV-nm}$, and temperature $T = 50\mathrm{mK}$ ~\cite{pan2021threeterminal,woods2021chargeimpurity,dassarma2023spectral}. Since the experimental contacts appear to be highly transparent, we choose the barrier voltage to be $V_{\text{barrier}}^{L/R}=0\, \hl{\text{meV}}$. 
Motivated by the experimental transport measurements in this configuration~\cite{microsoftquantum2023inasal}, the magnetic field $B$ was chosen to be $B = 1\,\text{T}$, oriented in the plane of the device and perpendicular to the nanowire. and the chemical potential over the range $\mu \in [-2\,\mathrm{meV}, 12\,\mathrm{meV}]$, using 400 steps. The latter range was determined using the experimental lever arm of $60\text{meV/V}$ together with the gate voltage range $-1.2\,\text{V}$ to $-1.0\,\text{V}$ where single channel transport was expected~\cite{microsoftquantum2023inasal}.

\begin{figure*}[ht]
    \centering
    \begin{subfigure}[b]{0.8\textwidth}\centering\includegraphics[width=\textwidth]{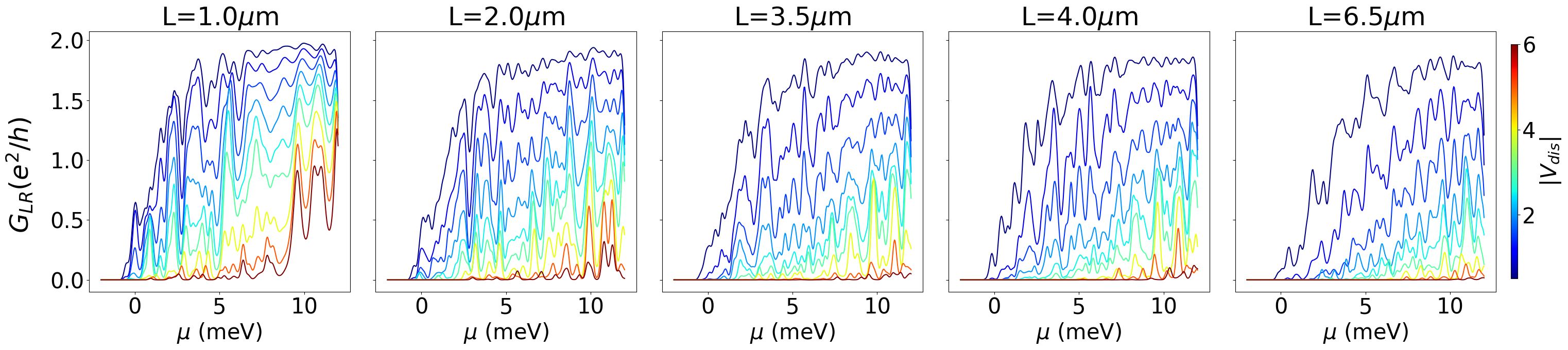}
        \caption{}
    \end{subfigure}
    \hfill
    \begin{subfigure}[b]{0.8\textwidth}
        \centering
        \includegraphics[width=\textwidth]{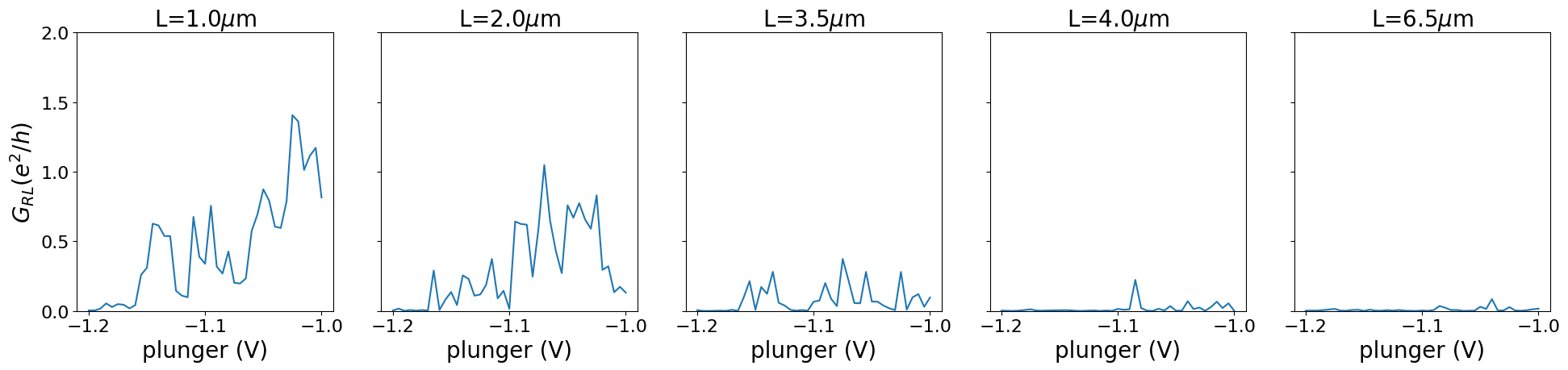}
        \caption{}
    \end{subfigure}
    \caption{ (a) Numerical non-local conductance $G_{LR}$ as a function of chemical potential $\mu$ (in meV) for different lengths $L$ and disorder strength $|V_{\hl{\text{dis}}}|$.
        The disorder strength of the different plots in the figure increases from blue to red while the length of the system increases from left to right. Each plot shows that the conductance vanishes below a critical chemical potential. At the same time, the critical chemical potential increases as the disorder strength and length increase. Note that to improve visibility, a chemical potential broadening of $0.14\,\textrm{meV}$ has been introduced to suppress mesoscopic fluctuations.
    (b) Experimental results for non-local conductance as a function of plunger voltages that are extracted from \JTadd{the 6th figure of} \cite{microsoftquantum2023inasal} for increasing lengths from left to right. As described in Sec.~\ref{sec:SC}, the range of plunger voltage corresponds to the chemical potential range plotted in the x-axis in panel (a). Comparing the profile of the conductance from panel (b) with the various disorders at panel (a), $|V_{\hl{\text{dis}}}|={4}\,\textrm{meV}$ appears to match the experiment the closest.}\label{fig:cond}
\end{figure*}

\begin{figure*}[ht]
    \centering
    \begin{subfigure}[b]{0.8\textwidth}
        \centering\includegraphics[width=\textwidth]{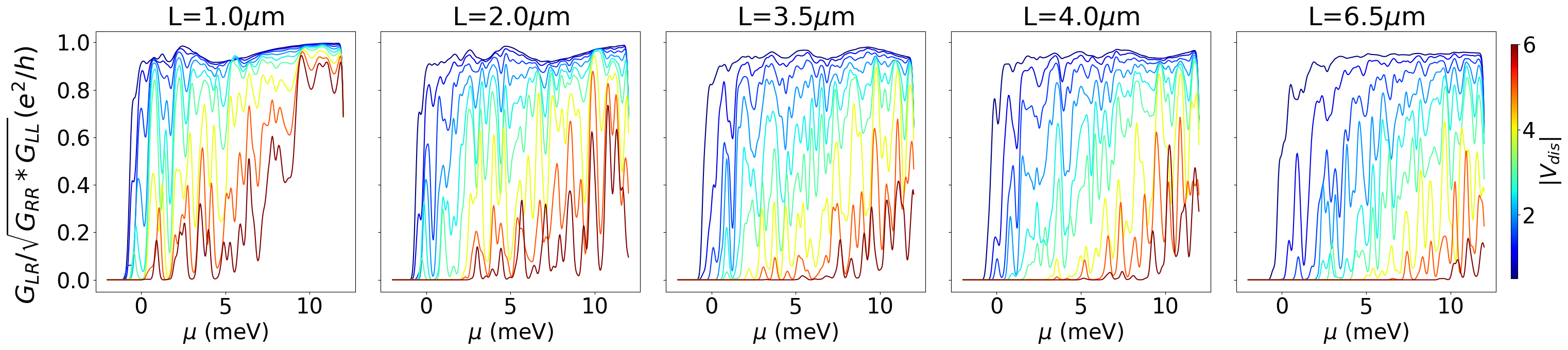}
        \caption{\comment{\HPcom{Hi Jacob, I noticed that you used `subfigure' and include the package `subcaption'; however i,t also makes the captions centered. Do you really have to use `subfigure' here? Can we combine the figure outside \LaTeX and include it as one figure?}}}
    \end{subfigure}
    \hfill
    \begin{subfigure}[b]{0.8\textwidth}
        \centering
        \includegraphics[width=\textwidth]{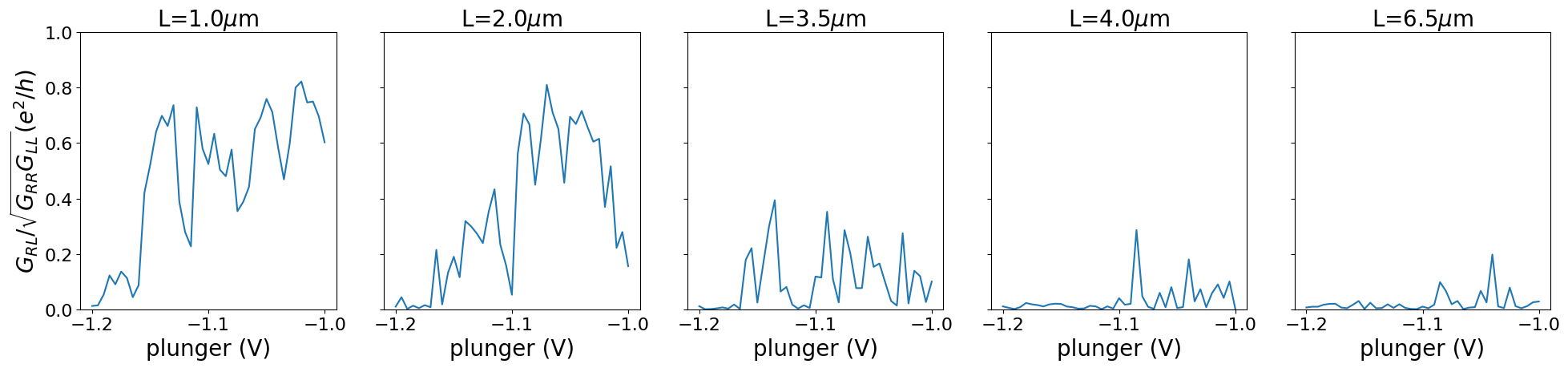}
        \caption{}
    \end{subfigure}
    \caption{ (a) Numerical normalized non-local conductance $G_{LR}/\sqrt{G_{RR}G_{LL}}$ as a function of chemical potential $\mu$ (in meV) for different lengths $L$ and disorder strength $|V_{\hl{\text{dis}}}|$ (meV).
        The disorder strength of the different plots in the figure increases from blue to red while the length of the system increases from left to right. Each plot shows that the conductance vanishes below a critical chemical potential. At the same time, the critical chemical potential increases as disorder strength and length increases. Note that to improve visibility, a chemical potential broadening of $0.14\,\textrm{meV}$ has been introduced to suppress mesoscopic fluctuations.
    (b) Experimental results for normalized conductance as a function of plunger voltages that are extracted from \JTadd{the 6th figure of} \cite{microsoftquantum2023inasal} for increasing lengths from left to right. As described in Sec.~\ref{sec:SC}, the range of plunger voltage corresponds to the chemical potential range plotted in the x-axis in panel (a). Comparing the profile of the conductance from panel (b) with the various disorders at panel (a), $|V_{\hl{\text{dis}}}|=\JTadd{4}\,\textrm{meV}$ appears to match the experiment the closest.}\label{fig:normcond}
\end{figure*}

The Majorana nanowire, in the above parameter regime and in the absence of disorder, has a gapless bulk with ballistic quasiparticles that would nominally lead to 
large non-local conductance $G_{LR}$ even for long wires. On the other hand, one can expect disorder to localize these quasiparticles and suppress the conductance at lengths 
exceeding the localization length~\cite{microsoftquantum2023inasal}. We include disorder in the Hamiltonian $H$ by choosing $V_{\hl{\text{dis}}}(x)$ to be Gaussian distributed with a correlation length $10\,\textrm{nm}$.
The non-local conductance results for Majorana nanowires for various lengths and disorder as well as the corresponding experimental results~\cite{microsoftquantum2023inasal} are shown in Fig.~\ref{fig:cond}.
While the actual disorder profile for the experimental results are unknown, for the theory simulations in Fig.~\ref{fig:cond}, we vary the rms magnitude of the disorder potential in the range $|V_{\hl{\text{dis}}}| \in [0.5\,\text{meV}, 5\,\text{meV}]$ in steps of $0.5\,\text{meV}$, 
while the wire length was varied as $L \in [1\,\mu\text{m}, 6.5\mu\text{m}]$. 
Mesoscopic conductance fluctuations make it difficult to directly compare the experimental data to the simulations. Generically, the non-local conductance in both the simulations and experiment vanish at the lower end of the chemical potential and increase as the chemical potential is increased. We therefore compare the non-local conductance averaged over the higher third of the range of chemical potential between the simulation and experiment shown in Fig.~\ref{fig:cond}.
Based on this comparison, we find that the best agreement between the simulations in Fig.~\ref{fig:cond}(a) and the experimental data~\cite{microsoftquantum2023inasal} reproduced in Fig.~\ref{fig:cond}(b) occurs around $|V_{\hl{\text{dis}}}| \sim 4\,\text{meV}$. { (Further error estimates and variance analysis can be found in Fig. \ref{fig:errors}.)} Simulations varying the correlation length of the disorder find that the correlation length chosen for Fig.~\ref{fig:cond} (i.e.  $l_{dis} = 10\ \text{nm}$) seems to fit 
the experiment best, though it should be noted that the actual disorder in the experiments~\cite{microsoftquantum2023inasal} likely involves a combination of disorder mechanisms with different characteristic lengths. For our purposes of qualitative comparison, we use a simplified model of assuming that one length-scale dominates.
It should be noted that the exact value of the estimated disorder depends on the criterion used to decide the best fit between theory and experiment.
However, the resulting uncertainty is within a factor of 2.
 Motivated by the experimental results for non-local conductance~\cite{microsoftquantum2023inasal} and to reduce the dependence on 
the contact resistance at the ends of the wire, we also compute the non-local conductance normalized by the local conductances i.e. $\frac{G_{LR}}{\sqrt{G_{RR}G_{LL}}}$,
where $G_{RL}$ and $G_{LR}$ are the non-local conductances, and $G_{RR}$ and $G_{LL}$ are the local conductances at each end.
The result for the normalized non-local conductance is shown in Fig.~\ref{fig:normcond} where we compare the simulation and experiment
in a way analogous to Fig.~\ref{fig:cond} and find that the disorder strength is consistent with the estimate from the normalized conductance.
Since the barrier transparency connecting the nanowire to the leads is high, it is not surprising that the normalization of the non-local conductance does not significantly 
modify the conclusion. Actually, the magnitude and length dependence of the local conductance is a complicated function of the details of the contact and will therefore 
not be discussed here. This is because the local conductances depend on a combination of Andreev reflection and normal reflection~\cite{blonder1982transition} which is less sensitive  to the disorder in the bulk of the nanowire. The contact effects can be expected to be relatively modest for longer wires where most of the suppression of conductance comes from bulk scattering. 
In contrast, the non-local conductance $G_{LR}$ plotted in Fig.~\ref{fig:cond} depends mostly on the transmission which is the most sensitive to the middle of the nanowire \hl{and} is the most relevant part for estimating nanowire disorder.

Note that the level of disorder (i.e. $|V_{\hl{\text{dis}}}|\sim 4\,\textrm{meV}$) estimated from the magnitude of non-local conductance is quite a bit larger than expected for the experimental device (see for example simulations~\cite{dassarma2023spectral,pan2024disordered}). This likely indicates that the multiple contacts present in this transport
device but not present in the qubit devices might have enhanced scattering.
Nonetheless, the results highlight the substantial uncertainty in the disorder present in experimental devices, and highlight the need for studying the in-situ parallel magnetic field characterization of Majorana nanowire devices. In addition, new techniques to better characterize and understand them may also help~\cite{taylor2024machine,taylor2025vision,pan2024disordered}.

\section{Conclusion}\label{sec:conclusion}
We have calculated the ballistic conductance in InAs nanowires as a function of disorder in order to understand the observable signatures of the helical gap arising from Zeeman splitting and spin-orbit coupling in Majorana platforms. We also calculate the conductance in the setup where the magnetic field is oriented perpendicular to the wire so that the helical gap does not exist, as motivated by a recent experimental study. Our main finding is that disorder has a profound effect on the conductance, and for disorder larger than the helical gap, the putative signatures of the gap (e.g., the re-entrant behavior of the conductance) are suppressed. We also find very strong Fabry-P\'erot resonances in the conductance, which appear complementary to the re-entrant behavior, implying that conductance measurement in InAs nanowires with the magnetic field oriented along the wire should manifest either the re-entrant behavior or strong resonances. Other parameters, in addition to disorder, strongly affecting (both qualitatively and quantitatively) the conductance are the SOC strength and the chemical potential mismatch between the two leads and the wire itself. Unfortunately, {many} of the key parameters (i.e., disorder, SOC strength, chemical potentials) are not quantitatively known independently in the SM-SC structures, making the situation difficult to figure out. In fact, even the Zeeman splitting is not well-known since the $g$-factor in the SM-SC nanowire cannot be measured in situ, {though this can be estimated from the gap closing magnetic field.}  These uncertainties actually make detailed and accurate conductance measurements in nanowires a necessity for experimental Majorana physics in SM-SC platforms since the helical gap provides an upper bound on the attainable topological gap in the system. It is therefore surprising and problematic that such ballistic conductance measurements in InAs nanowires (as presented in our Sec.~\ref{sec:SM}) have not been reported in the experimental literature, and the reported experiments in InSb~\cite{kammhuber2016conductance} \comment{[https://doi.org/10.1021/acs.nanolett.6b00051]}and PbTe~\cite{wang2023ballistic}\comment{[ https://doi.org/10.1021/acs.nanolett.3c03604]} nanowires, which are also utilized in SM-SC Majorana platforms, do not observe any signatures of the helical gap. Whether this arises from the weakness of the SOC strength or from the presence of large disorder strength or from some other mechanisms is a puzzle, which must be resolved for progress in the field.~\cite{sarma2025rashba}\comment{[cite the recent Rashba review by us]} {By contrast, a recent experiment does report ballistic conductance measurements in the actual InAs/Al hybrid platform with a magnetic field oriented normal to the SM and the SC, so that there is neither a helical gap nor any proximity superconductivity, but our results (presented in Sec.~\ref{sec:SC}) seem to indicate that the disorder in the sample may simply be too large. Since the spin-orbit coupling does not play a role in this measurement, the uncertainties with the spin-orbit coupling and $g$-factor become less of a problem for this measurement. However, the experimental device in this case, unlike the topological superconducting devices, had many contacts, which might have reduced the quality of the device. It should be noted, though, that both our theory and the experimental data have focused on one particular combination of the four conductance traces that are obtained in the experiment. This is a significantly more complicated scenario to interpret relative to normal state conductance, so we focus on qualitative comparison to simulations to make our conclusions.}  This again reinforces the importance of normal state conductance measurement in InAs nanowires in the presence of a magnetic field so that the basic helical band structure and the effective disorder strength can be figured out. If no helical gap exists in the normal state ballistic conductance, the topological superconducting { gap is weak}. Similarly, if the disorder strength is larger than the SC gap, there is no topological superconductivity. Detailed normal state ballistic conductance measurements in many different samples (presumably with different disorder)  as a function of various system parameters (e.g., gate voltage, tunnel barrier voltage, and applied magnetic field) should go a long way in resolving these questions. We see this as an urgent necessity at this point in the development of Majorana research in SM-SC hybrid platforms since the SOC is the essential key to creating Majorana-carrying topological superconductivity.

We note that, given the already large number of unknown system parameters in the problem, we have left out several physical effects that are straightforward to include in the theory if necessary, because of future experiments. For example, our theory is at $T=0$, and a finite temperature generalization is simple by including a Fermi function in the conductance calculation. Experimental base temperatures are typically $\sim 20$~mK (and even the electron temperature is likely $< 40$~mK), and the induced SC gap is $0.15$--$0.2$~meV $\sim 2$~K, and even the reported topological gap~\cite{microsoftquantum2023inasal}\comment{ [cite MSFT PRB]} is $\sim 30$--$60$~$\mu$eV $\approx 200$--$600$~mK, making any finite temperature Fermi function corrections to be exponentially small. The main effect of the temperature on the ballistic conductance would be to smoothen and soften the resonances, but such a smoothening would be weak in the context of the helical gap as long as the Zeeman splitting ($0.2$~meV $\sim 2$~K) is much larger than the temperature, which is the situation we consider (and also experimentally because of the very large InAs $g$-factor). Another relevant parameter is the detailed shape of the 1D confinement potential, which we have assumed to be a (parameter-free) sharp confinement since any other confinement model will introduce several unknown parameters in modeling the confinement, distracting from our main message of the manifestation of the helical gap on transport. Again, relaxing the sharp confinement model to a smooth adiabatic confinement potential is straightforward in the theory at the cost of several more unknown parameters, which must be varied, considerably complicating the results. Smoother confinement introduces adiabaticity, suppressing the sharp FP resonances (but depending crucially on the unknown parameters of the confinement model) -- for example, in the point contact geometry, where the leads are adiabatically transformed into a short 1D-like point contact, resonances are strongly suppressed.~\cite{he1989quantum,he1993quantum} \comment{[https://journals.aps.org/prb/abstract/10.1103/PhysRevB.40.3379;  https://journals.aps.org/prb/abstract/10.1103/PhysRevB.48.4629]}.  But the nanowires used in the SM-SC platforms are much closer to a sharply confined 1D channel than an almost zero-dimensional point contact, so the sharp confinement model, which perhaps overestimates the Fabry-P\'erot resonances, is likely to be a much more realistic model than adiabatic confinements relevant for point contacts.   We mention that both finite temperature and smooth potential effects have been considered for Rashba nanowires in a general context, and the qualitative conclusion is exactly what we describe, i.e., an overall (strongly parameter dependent) smoothening effect by both.~\cite{rainis2014conductance}\comment{[https://journals.aps.org/prb/abstract/10.1103/PhysRevB.90.235415]}
We also neglect interaction effects, which are negligible in InAs nanowires, and would have no quantitative effects on the ballistic conductance at the level of experimental interest-- interactions convert the wire into a helical Luttinger liquid, but any quantitative corrections of this Luttinger liquid are subtle and negligible in the results we present, where disorder effects dominate (in fact, disorder converts the Luttinger liquid generically to an 1D Anderson localized system in the thermodynamic limit which behaves as a ballistic conductor as long as the effective mean free path is longer than the wire length).

Finally, we mention that there has been one report in the experimental literature~\cite{quay2010observation}\comment{ [https://www.nature.com/articles/nphys1626]} of the `Observation of a one-dimensional spin-orbit gap in a quantum wire' where the ballistic conductance measurements were carried out in 1D Ge hole nanowires. Ge hole nanowires have completely different and very complex band structure compared with InAs electrons relevant for the SM-SC hybrid electronic Majorana platform, where both heavy holes and light holes are important and the $g$-factor is highly anisotropic (and in addition, strain effects are crucial)~\cite{laubscher2024majorana,laubscher2024germaniumbased}  \comment{[https://journals.aps.org/prb/abstract/10.1103/PhysRevB.109.035433;   https://journals.aps.org/prb/abstract/10.1103/PhysRevB.110.155431]}  In addition, the SOC coupling strength is typically very large in Ge holes and the disorder is typically very small, again in contrast to InAs electron nanowires under consideration in the current work. These Ge hole measurements do not observe any re-entrant conductance quantization, but only a dip in the conductance at a very high magnetic field (9T) which must produce a very large helical gap, particularly since the SOC coupling is also large in Ge holes. Although this hole experiment may be considered to be some evidence supporting some manifestation of the helical gap in the Ge hole nanowires at very high magnetic field, it cannot say anything about what happens for InAs electrons, which is the focus in the current work and in the experimental search for topological superconductivity in SM-SC hybrid platforms.

\hl{For completeness, we add a brief reminder of the standard clean-wire result: in the Rashba nanowire model with proximity-induced pairing, the topological phase transition is obtained from the closing and reopening of the bulk quasiparticle gap at $k=0$, which occurs at $V_Z^2=\Delta^2+\mu^2$ (topological for $V_Z^2>\Delta^2+\mu^2$; see Refs.~\cite{sau2010generic,sau2010nonabelian,lutchyn2010majorana,oreg2010helical}). Importantly, the visibility of a normal-state helical-gap signature in ballistic conductance is not a prerequisite for this criterion to be satisfied: for weak SOC and/or in the presence of disorder and Fabry--P\'erot resonances, the helical-gap conductance features can be strongly suppressed even though the system may still be in the topological regime, albeit with a parametrically small topological gap.}

We conclude by reiterating the importance of carrying out detailed measurements of ballistic conductance in InAs nanowires to search for the direct manifestation of the helical gap, which is an essential prerequisite for the existence of Majorana zero modes in these artificially engineered topological platforms.~\cite{sau2010generic,sau2010nonabelian,sarma2025rashba}\comment{ [cite Sau PRL 2010 and Sau PRN 2010, and our recent Rashba review here]} We also restate our two main conclusions from this work: (1) Non-observation of a  helical gap and the associated quantized conductance re-entrance as well as the absence of prominent resonances in the ballistic conductance by itself does not rule out the manifestation of topological superconductivity with emergent end-localized Majorana zero modes, but this most likely implies that the SOC (disorder) strength is small (large), indicating any emergent topological superconducting gap is weak (sec. \ref{sec:SM}); (2) a direct comparison of our ballistic conductance calculation in the presence of the superconductor with rotated magnetic field suppressing the gap with a recent experiment implies  a very large disorder in the system whose strength is larger than even the induced superconducting gap at zero magnetic field, most likely indicating that better experiments should be planned in the future with perhaps fewer contacts in order to estimate the in situ disorder operational in the experimental Majorana nanowires used for producing the topological qubits where additional contacts and leads are avoided as much as possible. \nocite{taylor_2026_18933032}

\section*{Acknowledgment}
This work is supported by the Laboratory for Physical Sciences (LPS) through the Condensed Matter Theory Center at Maryland. JRT and JDS thank the Joint Quantum Institute for additional support.  
HP is supported by US-ONR grant No.~N00014-23-1-2357.

\bibliography{Paper_SOC}
\appendix

\section{InAs without magnetic field and spin-orbit coupling}\label{app:InAs_no_field_no_SOC}
\begin{figure*}[ht]
    \centering
    \includegraphics[width=6.8in]{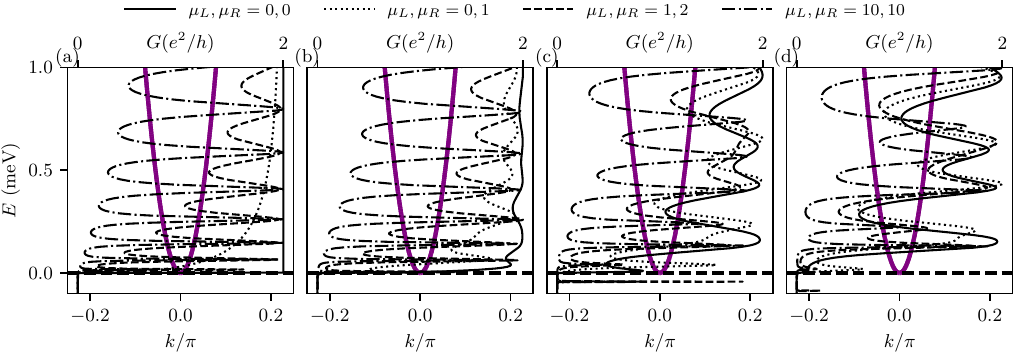}
    \caption{
        Tunneling conductance with various lead chemical potentials (black curves) and the \hl{spin-degenerate} band structure (purples curves) in a InAs nanowire with a zero spin-orbit coupling $\alpha=0$, wire chemical potential $\mu=0$ (horizontal dashed line), and a Zeeman field $V_Z=0$ without helical gap.
    The disorder strength increases from left to right: pristine wire in (a); $\sigma_\mu=0.05$ meV in (b); $\sigma_\mu=0.2$ meV in (c); $\sigma_\mu=0.3$ meV in (d). The conductance is for a single disorder realization without any ensemble averaging.
        }
    \label{fig:alpha0Vz0}

\end{figure*}

In this section, we present the band structure and the conductance properties of InAs nanowires without magnetic field and spin-orbit coupling, as shown in Fig.~\ref{fig:alpha0Vz0}.

\begin{figure*}[ht]
    \centering
    \begin{subfigure}[b]{0.45\textwidth}
        \centering\includegraphics[width=\textwidth]{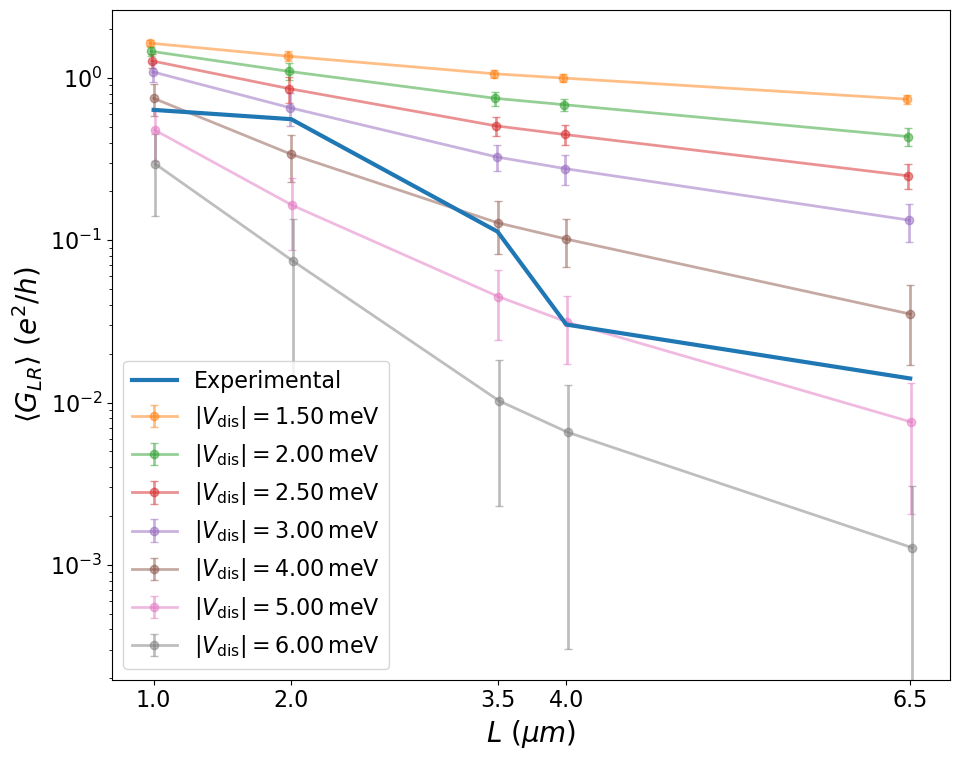}
        \caption{}
    \end{subfigure}
    \hfill
    \begin{subfigure}[b]{0.45\textwidth}
        \centering
        \includegraphics[width=\textwidth]{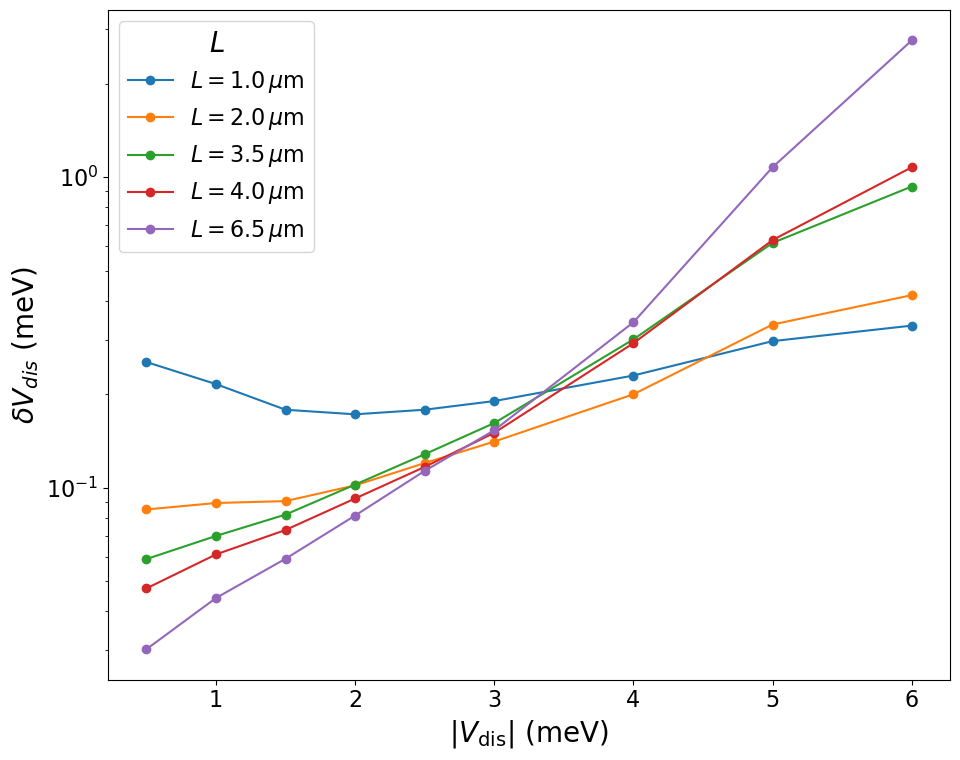}
        \caption{}
    \end{subfigure}
    \caption{\hl{(a) Diagram of $\langle G_{LR}\rangle$ conductance vs. L for different disorder magnitudes, along with experimental data from \cite{microsoftquantum2023inasal}. For simulated results $G_{LR}$ is averaged over $[5,10]$ meV, while for the experiment it is averaged over -1.1V to -1.015V. The experimental $G_{LR}$ seems to be consistent for a value of 4 meV. (b) Expected error $\delta V_\text{dis}$ of the disorder prediction $\delta V_{\text{dis}} = \frac{\sigma_G}{|\partial \langle G \rangle / \partial V_{\text{dis}}|}
$ where the total variance $\sigma_G$ is defined as $\sigma_G=\sqrt{(\sigma_G^{\text{th}})^2+(\sigma_G^{\text{exp}})^2}$ such that $\sigma_G^{\text{th}}$ and $\sigma_G^{\text{exp}}$ are the variance of $G_{LR}$ over $[5,10]$ meV and $[-1.1V,-1.015]$V for theory and experiment respectively. In both (a-b) 20 different simulated disorder realizations are used to average $G_{LR}$.}}\label{fig:errors}
\end{figure*}

\end{document}